\documentclass[11pt]{JHEP3}
\usepackage{graphicx}
\usepackage{epstopdf}
\usepackage{amsmath, amssymb}

\newcommand{\be}{\begin{eqnarray}}
\newcommand{\ee}{\end{eqnarray}}
\newcommand{\nn}{\nonumber}
\newcommand{\bn}{\begin{enumerate}}
\newcommand{\en}{\end{enumerate}}

\def\IC{\mathbb{C}}
\def\ID{\mathbb{D}}

\def\IP{\mathbb{P}}
\def\IR{\mathbb{R}}
\def\IZ{\mathbb{Z}}

\def\CI{{\cal I}}

\def\CN{{\cal N}}

\def\a{\alpha}
\def\b{\beta}
\def\g{\gamma}
\def\d{\delta}
\def\e{\epsilon}

\def\th{\theta}

\def\ch{\chi}
\def\w{\omega}

\def\D{\Delta}

\def\O{\Omega}

\def\half{\frac{1}{2}}
\def\thalf{{\textstyle \frac{1}{2}}}

\def\goto{\rightarrow}

\def\p{\partial}

\def\Tr{{\rm Tr}}
\def\tr{{\rm tr}}
\def\det{{\rm det}}

\def\da{{\dot{\a}}}
\def\db{{\dot{\b}}}
\def\dg{{\dot{\g}}}

\def\cN{\mathcal{N}}

\title{Superconformal Indices for Orbifold Chern-Simons Theories}

\author{Jaehyung Choi$^1$, Sangmin Lee$^2$ and Jaewon Song$^3$ \\
 $^1$Department of Physics and Astronomy, SUNY, Stony Brook, NY 11794-3800, USA \\
 $^2$Department of Physics \& Astronomy, Seoul National University, Seoul 151-747, Korea \\
 $^3$California Institute of Technology 452-48, Pasadena, CA 91125, USA
}

\abstract{We calculate the superconformal indices of recently 
discovered three-dimensional 
$\cN=4,5$ Chern-Simons-matter theories and compare them 
with the corresponding indices of supergravity on AdS$_4$ times orbifolds of $S^7$. 
We find perfect agreement in the large $N$ and large $k$ limit, 
provided that the twisted sector contributions at the fixed loci of the orbifolds are properly taken into account. 
We also discuss the index for the so-called ``dual ABJM'' proposal.}


\preprint{
CALT-68-2708
}

\dedicated{}

\begin{document}
\section{Introduction}
Recently, there has been a lot of progress in understanding the low energy effective theory of coincident M2-branes. 
Schwarz raised the possibility of using Chern-Simons theories to describe the superconformal theory on M2-branes \cite{Schwarz:2004yj}. The idea was first realized concretely by Bagger and Lambert \cite{Bagger:2006sk} and Gustavsson \cite{Gustavsson:2007vu} (BLG), where they constructed a Lagrangian for $SU(2) \times SU(2)$ Chern-Simons theory with $\cN=8$ supersymmetry. The BLG theory had some peculiarities that appeared puzzling from the M2-brane point of view. First, it was not clear how to extend this theory to describe arbitrary number of M2-branes. Second, a string/M-theoretic 
derivation of the BLG theory was lacking. 

Soon afterwards, the BLG theory was followed by a variety of superconformal Chern-Simons theories more clearly rooted in string/M-theory.\footnote{In this paper, we will focus 
on relatively new $\CN\ge4$ theories only. For a nice summary of more 
conventional $\CN \le 3$ Chern-Simons theories, 
see {\it e.g.} \cite{Gaiotto:2007qi} and references therein and thereto.} In a type IIB string theory setup, Gaiotto and Witten \cite{Gaiotto:2008sd} gave a general 
construction of $\CN=4$ Chern-Simons theories with one type of 
hyper-multiplets, where the theories were shown to be classified  
by an auxiliary Lie super-algebra. 
This construction was extended in \cite{HLLLP1} to include twisted hyper-multiplets, so that all $\CN \ge 4$ theories can, 
in principle, fit into the Gaiotto-Witten classification. 
It was also pointed out in \cite{HLLLP1} that the Gaiotto-Witten setup can be related via T-duality to M2-branes 
probing orbifold geometries. 

Aharony, Bergman, Jafferis and Maldacena \cite{ABJM} (ABJM) then performed an in-depth study of an especially simple and instructive case of $\CN=6$ theory 
with $U(N)\times U(N)$ gauge group. They gave convincing arguments that 
the Chern-Simons theory at level $k$ is dual to M-theory on 
AdS$_4 \times S^7/\IZ_k$. The ABJM theory thus opened up 
a laboratory for testing the long-missing AdS$_4$/CFT$_3$ duality 
and led to many exciting developments. 
More work on explicit construction and classification of $\CN=4,5,6$ theories can be found in \cite{Benna, Imamura:2008nn, HLLLP2,Bagger:2008se,Tera,Schnabl:2008wj,Imamura:2008dt,hohm}.

The aim of this paper is to compute the superconformal index \cite{Kinney:2005ej, Romelsberger:2005eg} 
for some $\CN=4,5$ quiver Chern-Simons theories constructed in 
\cite{HLLLP1,HLLLP2} and further studied in \cite{Benna, Imamura:2008nn,Tera}, which are believed to be dual to M-theory on AdS$_4$ times certain orbifolds of $S^7$.
The superconformal index, originally defined for 4-dimensional theories in \cite{Kinney:2005ej, Romelsberger:2005eg} counts the number of certain chiral operators. This index gets contributions only from short multiplets which cannot combine to form long multiplets as the parameters of the theory are varied. Just as for the Witten index \cite{Witten:1982df}, the superconformal index does not change under continuous change of parameters. The superconformal indices in 3, 5, 6 dimensions were constructed in \cite{Bhattacharya:2008zy, Dolan}, and the index for the ABJM theory has been computed in \cite{Bhattacharya:2008bja, Dolan}.

As in \cite{Bhattacharya:2008bja}, we compute the superconformal indices for $\cN=4, 5$ Chern-Simons theories in the large $N$ and large $k$ limit, where $N$ is the rank of the gauge group and $k$ is the Chern-Simons level. As explained in \cite{ABJM}, in the large $k$ limit, the circle fiber of $S^7$ shrinks. So, in our case, the compact space becomes an orbifold of $\IC \IP^3$. In the large $N$ limit with 
't Hooft coupling $\lambda = N/k$ fixed at a large value, the supergravity description is valid and we can compute the index by counting the Kaluza-Klein spectrum of the compact space. On the field theory side, since we have the effective coupling $\lambda$ which can be taken to zero, we can compute the index in the free theory limit. However, when the orbifold has a non-trivial fixed locus, we have massless twisted sector contributions which survive the large $N$ limit. This is an inherently stringy effect, so we cannot solely rely on supergravity description. Similar situation in four-dimensional quiver gauge theories has been discussed in \cite{Nakayama:2005mf}. We discuss how to figure out the twisted sector contribution to the index both in field theory and gravity side. After taking into account the twisted sector, we show that the indices in $\lambda = 0$ and $\lambda = \infty$ agree. 

The outline of this paper is as follows. In section 2, we briefly review the definition of the superconformal index, some salient features of the $\cN=4, 5$ theories we study in later sections, and the computation of the index in ABJM theory \cite{Bhattacharya:2008bja}. In section 3 and 4, we compute the index for $\cN=5$ and $\cN=4$ theories in field theory as well as in gravity. Perfect agreement is found in the $\CN=5$ without twisted sector contributions. Agreement is 
possible for $\CN=4$ theories if and only if the twisted sector 
contributions are taken into account. 
In subsection 4.3, we also briefly discuss the ``dual ABJM model'' proposed in \cite{Hanany:2008fj,Franco:2008um}. In section 5, we conclude with some open questions. Some useful formulas are provided in the appendix.  

\section{Reviews}

\subsection{Superconformal index in three dimensions}

In many supersymmetric theories, there exist short multiplets which contain smaller number of states than ordinary long multiplets. Some of these short multiplets can be combined into long multiplets as the parameters of the theory are varied, but there are others that cannot. The superconformal index defined in \cite{Kinney:2005ej,Romelsberger:2005eg} receives contributions only from the latter states, so it is kept constant while the couplings of the theory change. 

In three dimensions, the bosonic subgroup of the superconformal group $OSp(\cN |4)$ is $SO(3, 2) \times SO(\cN)$, and its maximally compact subgroup is $SO(2)\times SO(3) \times SO(\CN)$, where $\CN$ is the number of supersymmetries. 
Following \cite{Bhattacharya:2008zy,Bhattacharya:2008bja}, we denote the eigenvalues of the Cartan generators of $SO(2)\times SO(3)\times SO(\CN)$ by $\e_0$, $j$ and $h_i$ $(i=1,\cdots, [\CN/2])$. 
In radial quantization (compactifying the theory on $\IR \times S^2$), $\e_0$ and $j$ are interpreted as energy and angular momentum. 
There are $4\cN$ real supercharges with $\e_0=\pm 1/2$. 

In this paper, we will compute the three-dimensional version of the index of \cite{Kinney:2005ej}, 
\be \label{WIndex}
 \CI(x,\{y_i\}) = \Tr \left[(-1)^F x^{\e_0+j} y_1^{h_2} \cdots y_{[\cN/2]-1}^{h_{[\cN/2]}}\right]\,.
\ee
The definition of the index singles out a particular supercharge $Q$ with charges $\e_0=+1/2$, $j = -1/2$, $h_1 = 1$, $h_i = 0$ $(i\ge2)$. The superconformal algebra implies 
that 
\be
 \{Q, Q^\dagger \} = \Delta \equiv  \epsilon_0 - j -h_1 \,.
\ee
The short multiplets contributing to the index satisfy $\Delta = 0$ and 
are annihilated by both $Q$ and $Q^\dagger$.  Note that these states can be 
interpreted as elements of $Q$-cohomology class. 
Since $\Delta = Q Q^\dagger + Q^\dagger Q$, we can think of $Q$ as 
analogous to the $d$ operator in de Rham cohomology and $\Delta$ to the Laplacian operator. 

On the field theory side, the index in the free theory limit can be computed using a matrix-integral formula \cite{Aharony:2003sx},  
\be
\CI(x,\{y_i\}) = \int \prod_a DU_a \exp \left(\sum_R \sum_{n=1}^\infty \frac{1}{n} f_R(x^n , y_1^n,\cdots y_{[\cN/2]}^n) \ch_{R} (\{U_a^n\})\right) \,,
\label{ind-ft}
\ee
where $R$ denotes the representations of matter fields (``letters''), 
$f_R$ is the index computed over the single letters without restriction 
on gauge invariance, and $\chi_R$ is the group character. In the case of $U(N)$ gauge theory with bi-fundamental matter fields,  
\be
\ch_{ab} (\{U_c\}) = \Tr U_a \Tr U_b^{\dagger} \,.
\ee
The index (\ref{ind-ft}) enumerates all possible gauge-invariant multi-trace operators. 
The standard rule of Bose statistics (or equivalently the plethystic integral) 
relates the ``multi-particle'' index $\CI$ to the ``single-particle'' index $I_{\rm sp}$ through 
\be
\CI(x,\{y_i\}) = \exp \left( \sum_{k=1}^{\infty} \frac{1}{k} I_{\rm sp} (x^k,\{y^k_i\}) \right) \,.
\ee

On the gravity side, the single particle index $I_{\rm sp}$ can be computed from  the Kaluza-Klein spectrum of eleven-dimensional supergravity compactified on the internal 
seven-manifold. For $S^7$, $I_{\rm sp}$ was computed in \cite{Bhattacharya:2008zy} using the known Kaluza-Klein spectrum in supergravity \cite{Gunaydin:1985tc,Biran:1983iy}. For orbifolds of $S^7$, the index counts 
states in untwisted and twisted sectors. The untwisted sector 
is simply the subset of the spectrum for $S^7$ that is 
invariant under the orbifold action. 
If the orbifold has fixed points, there may be additional twisted sector states 
localized at the fixed points. We will see that $\CN=5,6$ orbifolds 
have no twisted sector contributions, while $\CN=4$ orbifolds 
do have such contributions. 

As usual, the index should not change as we vary the continuous parameter of the theory \cite{Witten:1982df}. Contrary to the case of Yang-Mills theory, the effective coupling of the Chern-Simons theory $\lambda = N/k$ is discrete. But, when we take the large $N$ limit, $\delta \lambda = - \lambda^2 /N $ becomes effectively continuous. So we can take the limit of 't Hooft coupling $\lambda = N/k$ to be zero, and then the theory becomes free. This means that we also have to take the Chern-Simons level $k$ to be infinite, so $k$ never enters into our calculation.
In the following discussion, we take both large $N$ and large $k$ limit.

%
%

%
%
\subsection{$\cN=4,5,6$ Chern-Simons matter theories}
In this section, we give a short summary of the $\cN = 4,5,6$ superconformal Chern-Simons matter theories we will study in later sections. Our notations closely follow those of \cite{HLLLP1,HLLLP2}. 
Our discussion will be brief, and we refer the reader to the 
original papers \cite{Gaiotto:2008sd, HLLLP1,ABJM,Benna, Imamura:2008nn, HLLLP2,Bagger:2008se,Tera,Schnabl:2008wj,Imamura:2008dt} for details.
The field theory computations in this paper will be done 
in free theory limit, 
so for most purposes, it is sufficient to recall the matter content 
and gauge symmetry of the theory. We proceed in descending order 
of number of supersymmetries. 

\paragraph{$\CN=6$ ABJM theory}

The gauge group is $U(M)\times U(N)$. The theory makes 
sense at quantum level if and only if the rank of the gauge group and the Chern-Simons level $k$ satisfy $|M-N| \le k$ \cite{Aharony:2008gk}. The matter content is summarized by the following table: 
\be
\begin{array}{c|cccc}
  & \; \Phi_\a \;  & \;\bar{\Phi}^\a \; & \;\Psi^{\a}\; & \;\bar{\Psi}_{\a} \; \\ \hline
U(M)\times U(N) & (M,\bar{N}) & (N,\bar{M}) & (M,\bar{N}) & (N,\bar{M}) \\
SO(6)_R  & \mathbf{4} & \mathbf{\bar{4}} & \mathbf{\bar{4}} &
\mathbf{4}
\end{array}
\ee
To compare with the index computation of \cite{Bhattacharya:2008bja}, we choose the convention for $SO(6)_R$ highest 
weights such that $\mathbf{4}$ representation have $(h_1,h_2,h_3,h_4) = (+\half,+\half,+\half,-\half)$.

As explained in \cite{ABJM}, 
the moduli space of vacua of this theory is 
(a symmetric product of) $\IC^4/\IZ_k$. 
The scalars $\Phi_\a$ can be thought of as coordinates on $\IC^4$. 
The orbifold action $\IZ_k$, a residual discrete gauge symmetry 
on the Coulomb branch, acts on $\Phi_\a$ as $\Phi_\a \goto e^{2\pi i /k} \Phi_\a$.

\paragraph{$\CN=5$ theory}

The gauge group is $O(M)\times Sp(2N)$.\footnote{We use the notation in which $Sp(2N)$ has rank $N$.}
Let us denote the generators of $O(M)$ and $Sp(2M)$ as
$M_{ab}$ and $M_{\dot a\dot b}$, respectively.
The invariant anti-symmetric tensor of $Sp(2M)$ is denoted by
$\omega_{\dot a\dot b}$. The invariant tensor of $Sp(4)=SO(5)_R$ is denoted by $C_{\a\b}$.
We denote the bi-fundamental matter fields 
by
\begin{equation}
 \Phi_\alpha^{a\dot a}\,, ~~~
 \Psi_\alpha^{a\dot a} \,.
\end{equation}
They obey the reality condition of the form
\begin{equation}
 \bar\Phi^\alpha_{\dot aa} ~=~
 (\Phi_\alpha^{a\dot a})^\dagger ~=~
 \delta_{ab}\omega_{\dot a\dot b}C^{\alpha\beta}\Phi_\beta^{b\dot b},
 \label{real5}
\end{equation}
and similarly for the fermions.

As explained in \cite{HLLLP2}, 
the moduli space of vacua of this theory is $\IC^4/\ID_{k}$ 
where $\ID_{k}$ is the binary dihedral group with $4k$ elements. 
The dihedral group is generated by 
\be
\a \; : \; \Phi_\a \;\; \goto \;\; e^{\pi i/k} \Phi_\a \,,
\;\;\;\;\;
\b \; : \; \Phi_\a \;\; \goto \;\; C_{\a\b} \bar{\Phi}^\b \,.
\ee

\paragraph{$\CN=4$ quiver theories}

We use $(\a,\b ; \da,\db)$ doublet indices
for the $SU(2)_L\times SU(2)_R$ $R$-symmetry group. 
We denote the invariant tensors by $\epsilon_{\alpha\beta},
\epsilon_{\dot{\a}\dot{\b}}$ and their inverses by $\epsilon^{\alpha\beta},
\epsilon^{\dot{\a}\dot{\b}}$ such that $\epsilon^{\a\g}\epsilon_{\g\b}=
\delta^\a_{\ \b}$, $\epsilon^{\da\dg}\epsilon_{\dg\db}=
\delta^\da_{\ \db}$.
The hyper-multiplets are denoted by $(q_\a, \psi_\da)$ and 
the twisted hyper-multiplets by $(\tilde{q}_\da,\tilde{\psi}_\a)$. 
A doublet of $SU(2)_L$ has the $SO(4)$ 
highest weight $(h_1,h_2)=  (\half,\half)$ and 
a doublet of $SU(2)_R$ has $(\half,-\half)$. 

\begin{figure}[htbp]
\begin{center}
\includegraphics[width=10.5cm]{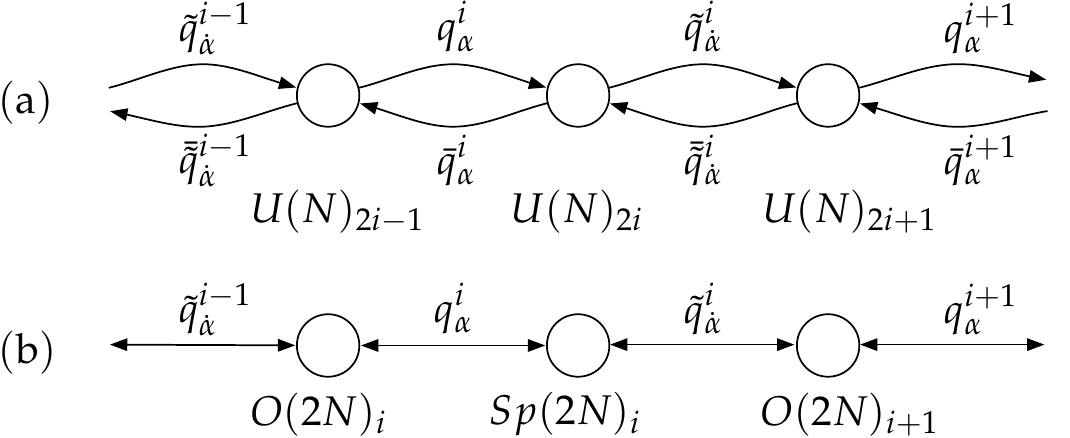}
\caption{Matter content of (a) $U$-type and (b) $OSp$-type quiver theories. } 
\label{quiver}
\end{center}
\end{figure}

We will consider two types of $\CN=4$ quiver gauge theories : 
the $U(M|N)$-type and the $OSp(M|N)$-type \cite{HLLLP1} 
or more briefly $U$-type and $OSp$-type. 
The $U$-type quivers can be viewed as an extension of the $\cN=6$ theory. It consists of a product of $2m$ $U(N_i)$ gauge groups. The ranks can be different in general, but for simplicity 
we will assume that they are all equal, so the gauge group is $U(N)^{2m}$. There are $m$ hypers $(q^i_\a, \psi^i_\da)$ in $(N,\bar{N})$ of $U(N)_{2i-1}\times U(N)_{2i}$ and $m$ twisted hypers $(\tilde{q}^i_\da, \tilde{\psi}^i_\a)$ in $(N,\bar{N})$ of $U(N)_{2i}\times U(N)_{2i+1}$. 
The hermitian conjugates ($\bar{q}_\a = \e_{\a\b} (q^\dagger)^\b$, etc.) 
belong to the same $R$-symmetry representation, 
but form anti-bi-fundamental representations under the gauge groups. 
The matter content of the $U(M|N)$-type quiver theory is summarized in Fig. \ref{quiver}(a). 

The $OSp$-type quivers can be regarded as 
an extension of the $\cN=5$ theory. The quiver consists of an alternating series of $m$ factors of $O(M_i)$ and $Sp(N_i)$. 
For simplicity, we assume that the gauge group is $[O(2N)\times Sp(2N)]^m$. There are $m$ hypers $(q^i_\a, \psi^i_\da)$ in bi-fundamental of $O(2N)_{i}\times Sp(2N)_{i}$ and $m$ twisted hypers $(\tilde{q}^i_\da, \tilde{\psi}^i_\a)$ in bi-fundamentals of $Sp(2N)_{i}\times O(2N)_{i+1}$. 
The hermitian conjugates obey a reality condition similar to  (\ref{real5}) with $C^{\a\b}$ replaced by $\e^{\a\b}$ or $\e^{\da\db}$.
The matter content of the $OSp$-type quiver theory is summarized in Fig. \ref{quiver}(b). 

The moduli space of vacua of the $U$-type theories were 
studied in \cite{Benna} and \cite{Tera}. 
When the Chern-Simons level $k$ is unity, the solution to F-term and D-term conditions gives the moduli space of vacua 
$(\IC^2/\IZ_m)^2$, where the generators of $\IZ_m$ act on coordinates of $\IC^4$ as 
\be
(z_1,z_2 , z_3,z_4) \sim (\w_m z_1, \w_m z_2 ,  z_3,z_4) 
\sim ( z_1, z_2 ,  \w_m z_3, \w_m z_4) 
\;\;\; (\w_m \equiv e^{2\pi i/m}) \,.
\label{orbi1}
\ee
For $k>1$, as in the $\CN=6$ theory, the residual discrete gauge symmetry induces further orbifolding by 
\be
(z_1,z_2 , z_3,z_4) \sim 
\w_k ( z_1, z_2 ,z_3,  z_4) \,.
\label{orbi2}
\ee
Note that all the orbifold actions (\ref{orbi1}, \ref{orbi2}) can 
be generated by two (as opposed to three) generators. 
In the simple case where $m$ and $k$ are relatively prime, we may choose the independent generators to be 
\be 
\label{orbiaction}
(\w_m, \w_m , 1, 1) \;\;\; {\rm and} \;\;\; 
(\w_{mk} , \w_{mk} ,  \w_{mk} ,\w_{mk}) \, ,
\ee 
and say that the moduli space is $\IC^4/(\IZ_m \times \IZ_{mk})$.
Similarly, the moduli space of the $OSp$-type theories can be shown to be  $(\IC^2/\IZ_m)^2 /\ID_k \simeq \IC^4/(\IZ_m \times \ID_{mk})$. The orbifold action $\IZ_m$ is the same as (\ref{orbi1}) and the action of $\ID_k$ is the same 
as in the $\CN=5$ theory.

For later purposes, let us review some details of the supersymmetry algebra of the $\cN=4$ theories. We denote the supercharges by $Q_{\a\da}$ and write their components as $Q_{\pm\pm}$. 
The matter fields can be written as $q_{\pm}$, $\psi_{\pm}$ and so on. 
In this notation, 
the special supercharge involved in the definition of 
the superconformal index is $Q_{++}$. 
The supersymmetry transformation rule includes 
\be
\left[ Q_{++}, q_+^i \right] = 0 \,, 
\;\;\;\;\; 
\left[ Q_{-+}, q_+^i \right] = \psi_+^i \,, 
\;\;\;\;\; 
\left[ Q_{++}, \tilde{\psi}_+^i \right] =\tilde{q}_+^i q_+^{i+1} \bar{q}_+^{i+1}  - \bar{q}_+^i q_+^i \tilde{q}_+^i 
\,.
\label{n4susy}
\ee
and similar relations obtained by hermitian conjugation and/or 
exchange of hypers with twisted hypers.

\subsection{Indices for $\CN=6$ ABJM theories}

We now review the computation of the index of the $\CN=6$ theory 
following \cite{Bhattacharya:2008bja}.
\footnote{
The same index was computed in \cite{Dolan} using a different method of evaluating the integral (\ref{n6-int}). 
The same reference also gives general formulas for 
the index for superconformal algebra $OSp(2N|4)$. 
See also a related work \cite{Dolan2} where the indices of $SO/Sp$ gauge theories (in four dimensions) were computed, 
which complements the method we develop in the next section.} 
We will be slightly more general and allow the gauge group to be $U(M) \times U(N)$. The compact bosonic subgroup of the superconformal group is $SO(2) \times SO(3) \times SO(6)$, 
and the index is given by
\be
\CI(x,y_1,y_2) = \Tr \left((-1)^F x^{\e_0 + j} y_1^{h_2} y_2^{h_3} \right) \,,
\ee
where $h_{i}$'s are the second and third Cartan charges of $SO(6)$.
The single letter index for bi-fundamental and anti-bi-fundamental matter fields are
\be
\label{lett-6a}
f_{12} = \frac{x^{1/2}}{1-x^2}\left(\sqrt{\frac{y_1}{y_2}} + \sqrt{\frac{y_2}{y_1}}\right) - \frac{x^{3/2}}{1-x^2}\left(\sqrt{y_1y_2} + \frac{1}{\sqrt{y_1y_2}}\right) \,,
\\
\label{lett-6b}
f_{21} = \frac{x^{1/2}}{1-x^2}\left(\sqrt{y_1y_2} + \frac{1}{\sqrt{y_1y_2}}\right) - \frac{x^{3/2}}{1-x^2}\left(\sqrt{\frac{y_1}{y_2}} + \sqrt{\frac{y_2}{y_1}}\right) \,.
\ee		
On the field theory side, the superconformal index is given by 
\be
\label{n6-int}
\CI = \int DU_1DU_2 \exp \left(\sum_{a,b} \sum_{n=1}^\infty \frac{1}{n} f_{ab}(x^n , y_1^n, y_2^n)  \Tr(U_a^n) \Tr(U_b^{\dagger n})\right) \,.
\ee
The only difference from \cite{Bhattacharya:2008bja} is that now $U_2$ is an element of $U(M)$. Now, we make the change of variables 
in a standard way in large $N$ computations,  
\be
\rho_n = \frac{1}{N} \Tr U_1^n, \;\;\; \chi_n = \frac{1}{M} \Tr U_2^n \,.
\label{rhoch}
\ee 
Then the measure is given by (a derivation of this measure factor is given in appendix \ref{meas})
\be
 DU_1 = \prod_{n=1}^{\infty} d\rho_n \exp \left( -N^2 \sum_n \frac{\rho_n \rho_{-n}}{n} \right ) \,,
 \\
 DU_2 = \prod_{n=1}^{\infty} d\chi_n \exp \left(-M^2 \sum_n \frac{\chi_n \chi_{-n}}{n} \right ) \,.
\ee
Substituting this, and writing $M = N + m$, we get 
\be
\CI = \prod_n d\rho_n d\chi_n \exp \left(-\frac{N^2}{2} \sum_n \frac{1}{n} C_n^T M_n C_n \right) \,.
\ee 
Here, $C_n^T =(\chi_n ~ \rho_n ~\chi_{-n} ~ \rho_{-n} ) $ and
\be
 M_n = \left( \begin{array}{cccc}
  0 & 0 & (1+\alpha)^2 & -(1+\alpha) f_{21;n} \\
  0 & 0 & -(1+\alpha) f_{12;n} & 1 \\
  (1+\alpha)^2 & - (1+\alpha) f_{12;n}  & 0 & 0 \\
  -(1 +\alpha) f_{21;n} &1 & 0 & 0 
 \end{array} \right) \,,
 \label{m144}
\ee
with $\alpha = m/N$ and $f_{ab;n} \equiv f_{ab}(x^n,y_1^n,y_2^n)$. 
The overall normalization of the Gaussian integral is fixed by requiring that $\CI(x=0,y_i=0)=1$. The final result is
\be
 \CI &=& \prod_n \frac{(1+\alpha)^2}{\sqrt{\det M_n}} 
 \;=\; \prod_n \frac{(1-x^{2n})^2}{(1 - \frac{x^n}{y_1^n})(1 - \frac{x^n}{y_2^n})(1 - x^n y_1^n)(1 - x^n y_2^n)} \,, 
 \nn \\
 I_{\rm sp} &=& \frac{x}{y_1-x}+\frac{1}{1-x y_1}+\frac{x}{y_2-x}+\frac{1}{1-x y_2}- \frac{2}{1-x^2} \,.
\ee
Note that the result is independent of $\a$ as we take the large $N$ limit.
This agrees with the observations on the gravity dual \cite{Aharony:2008gk}. The only difference comes from the torsion flux or in type IIA description, NS-NS 2-form flux through $\IC\IP^1 \subset \IC\IP^3$. Since the torsion does not affect the classical equations of motion, we expect exactly the same set of graviton states as in $U(N) \times U(N)$ theory. The torsion flux makes difference in the baryonic spectrum, but they do not contribute to the index in this limit.  In \cite{Aharony:2008gk}, it was also shown that this theory is superconformal only if $|M-N| \le k$. But in our field theory calculation, we are taking limit of $\lambda = N/k \to 0$. Therefore, we do not expect any sign of inconsistency for $|M-N|$ large. 
In view of these observations, we will neglect the differences 
in the ranks of the gauge group in the following sections.

\newpage

\section{Indices for $\CN=5$ theories}
In this section, we calculate the index of the $\cN=5$ $O(2N) \times Sp(2N)$ theory of \cite{HLLLP2}. 
The bosonic subgroup of the superconformal group is 
$ SO(3, 2) \times SO(5) $, so we define the index to be
\be
 \CI = \Tr \left[ (-1)^F x^{\epsilon_0 + j} y^{h_2} \right] \,.
\ee
where $h_2$ is the second Cartan charge of $SO(5)$. 

\subsection{Field theory}

We begin with the formula for the index of a free field theory 
with two gauge groups and bi-fundamental matter fields only, 
\be \label{MIntSpO}
\CI = \int DA DB \exp \left( \sum_{n=1}^{\infty} \frac{1}{n} f(x^n, y^n) \Tr(A^n) \Tr(B^{n}) \right)\,,
\ee
For the $\CN=5$  theory, we take $A \in SO(M)$ and $B \in Sp(2N)$. 
The function $f$ denotes the index of $\D=0$ ``letters''. 
The letters contributing to $f$ are summarized in the following table: 
\begin{table}[htbp]
\begin{center}
\begin{tabular}{|c|c|c|c|c|}
\hline
type & operators & $\e_0$ & $j$ & $SO(5)$ \\
\hline
bosons & $\Phi$ & $\thalf$ & $0$ &  $(\thalf,\thalf)$ \\
fermions & $\Psi$ & $1$ & $\thalf$ &  $(\thalf,\thalf)$ \\
\hline 
derivatives & $\partial$ & $1$ & $1$ & $(0,0)$ \\
\hline 
\end{tabular}
\end{center}
\caption{The $\D=0$ letters of the $\CN=5$ theory.}
\label{tb:osp-letter}
\end{table}

Contrary to the $\cN=6$ case, there is no distinction between 
bi-fundamental and anti-bi-fundamental matter fields. 
We also lose a Cartan generator and, consequently, 
the $\mathbf{4}$ and $\mathbf{\bar{4}}$ representations 
of $SO(6)$ become the same representation of $SO(5)$. 
It follows that, to obtain the index over the letters of the $\CN=5$ theory, 
we can simply set $y_2=1$ in the $\CN=6$ result (\ref{lett-6a}) or (\ref{lett-6b}). 
The result is 
\be
f(x, y) \;=\; \frac{ x^\half}{1-x^2} \left( \sqrt{y}+ \frac{1}{\sqrt{y}} \right) - \frac{x^{\frac{3}{2}}}{1-x^2} \left( \sqrt{y}+ \frac{1}{\sqrt{y}}  \right) 
\;=\; \frac{x^\half}{1+x} \left( \sqrt{y}+ \frac{1}{\sqrt{y}}  \right) \,.
\ee

\subsubsection*{Integration measure}

\paragraph{$SO(2N)$ :} 
Any element $A \in SO(2N)$ can be block-diagonalized into the form 
\be
A = \bigoplus_{i=1}^{N}
\begin{pmatrix}
\cos \a_i & -\sin \a_i \\
\sin \a_i & \cos \a_i 
\end{pmatrix} 
\;\;\;\;\; (|\a_i| \le \pi) \,.
\ee 
In this basis, 
the Harr measure over the $SO(2N)$ group manifold is (see appendix A) 
\be
DA = \prod_{i=1}^N d\alpha_i \prod_{i<j} \sin^2\left(\frac{\a_i-\a_j}{2}\right)\sin^2\left(\frac{\a_i+\a_j}{2}\right)\,. 
\ee
{}Throughout this subsection, we will suppress unimportant overall normalization constants. 
In the large $N$ limit, we can introduce the eigenvalue distribution function 
\be
 \rho(\theta) = \sum_i \delta (\theta - \alpha_i).
\ee 
Since the ``eigenvalues'' of $SO(2N)$ come in pairs $(e^{\pm\a_i})$, 
we can choose $\alpha_i > 0$ without loss of generality and restrict the domain of $\rho(\theta)$ to be $[0, \pi]$, so that 
\be
 \int_{0}^{\pi} \rho(\theta) d \theta = N.
\ee
Now, instead of integrating over $\alpha_i$'s we can integrate over the Fourier modes of $\rho$ : 
\be
\rho(\theta) = \frac{1}{\pi} \left[ \half \rho_0+ \sum_{n\ge1} \rho_n \cos(n\theta) \right]\,, 
\;\;\;\;\;
\rho_n = 2 \int_0^\pi \rho(\theta) \cos(n\theta) d\theta \,.
\ee
The normalization of $\rho_n$ is chosen such that\footnote{The normalization here differs from that in (\ref{rhoch}) 
by a factor of $N$. This is to emphasize that the finite shift in (\ref{so2nf}) survives the large $N$ limit. } 
\be
 \Tr(A^n) = 2 \sum_i \cos (n\a_i) = 2 \int d\theta \rho(\theta) \cos (n \theta) = \rho_n \,.
\ee
We can rewrite the measure factor as 
\be
 DA &=& \prod_{i} d\alpha_i\exp \left( \sum_{i < j} \left( \log \left| \sin^2 \left(\frac{\alpha_i - \alpha_j}{2} \right) \right| + \log \left| \sin^2 \left(\frac{\alpha_i + \alpha_j}{2} \right) \right| \right) \right) 
 \nonumber \\
  &=& \prod_{i} d\alpha_i\exp \left( -2 \sum_{n=1}^{\infty} \sum_{i < j} \left( \frac{\cos(n (\alpha_i - \alpha_j) )}{n} + \frac{\cos(n(\alpha_i+\alpha_j))}{n}\right) \right) 
  \\
  &=& \prod_{i} d\alpha_i\exp \left( - \sum_{n=1}^{\infty} 
  \left\{ \sum_{i, j} \left( \frac{\cos(n (\alpha_i - \alpha_j) )}{n} + \frac{\cos(n(\alpha_i+\alpha_j))}{n}\right) - \sum_i \frac{\cos(2n\a_i)}{n} \right\}\right) \,.
  \nn
\ee
In the large $N$ limit, 
\be
&&\sum_{i, j} \left( \frac{\cos(n (\alpha_i - \alpha_j) )}{n} + \frac{\cos(n(\alpha_i+\alpha_j))}{n}\right)
\nn \\
&\goto& 
\int d\theta_1 d\theta_2  \rho(\theta_1) \rho(\theta_2) \left( \frac{\cos(n (\theta_1 - \theta_2) )}{n} + \frac{\cos(n(\theta_1 + \theta_2)}{n}\right)   
  \;=\; \frac{1}{2n} \rho_{n}^2  \,,
\\
&&\sum_i \frac{\cos(2n\a_i)}{n}  
\; \goto \; 
\int d\th \rho(\th) \frac{\cos(2n\th)}{n} = \frac{1}{2n} \rho_{2n} \,.
\ee
To summarize, the integration measure of $SO(2N)$ is 
\be
DA = \prod_m d\rho_m \exp\left( - \sum_{n\; {\rm odd}} \frac{1}{2n} \rho_n^2  
- \sum_{n\;{\rm even}} \frac{1}{2n} (\rho_n -1)^2\right) \,.
\label{so2nf}
\ee 
This measure is consistent with the fact that 
\be
\langle \tr A^{2k+1} \rangle = 0, 
\;\;\;\; 
\langle \tr A^{2k} \rangle = 1,
\ee
which holds exactly (without large $N$ approximation) 
for $2k < N$. 

\paragraph{$SO(2N+1)$ :} 
The discussion proceeds in parallel with that of $SO(2N)$ 
with two important differences. First, the Harr measure over 
the $SO(2N+1)$ group manifold is given by
\be
DA = \prod_{i=1}^N d\alpha_i \prod_{i<j} \sin^2\left(\frac{\a_i-\a_j}{2}\right)\sin^2\left(\frac{\a_i+\a_j}{2} \right) \prod_{i} \sin^2\left(\frac{\a_i}{2}\right)
\,.
\ee
The last factor $\prod_i \sin^2(\a_i/2)$ induces a linear term in $\rho_n$ for every (even and odd) $n$. However, this shift is precisely compensated 
by another shift in the definition of $\rho_n$,  
\be
\rho_n \equiv \tr(A^n) = 2 \sum_i \cos(n \a_i) +1 \,.
\ee
As a result, the final form of the measure of $SO(2N+1)$ 
in the large $N$ limit is identical to that of $SO(2N)$ in (\ref{so2nf}).

\paragraph{$Sp(2N)$ :}

The Harr measure over the $Sp(2N)$ group manifold is given by
\be
DB = \prod_{i=1}^N d\beta_i \prod_{i<j} \sin^2\left(\frac{\b_i-\b_j}{2}\right)\sin^2\left(\frac{\b_i+\b_j}{2} \right) \prod_{i} \sin^2\b_i 
,.
\ee
Compared to the $SO(2N)$ case, the $\prod_i \sin^2\b_i$ term 
over-compensates the shifts for $n=2k$, resulting in the net shift
$+1$ as opposed to $-1$ of $SO(2N)$. In other words, with the definition, 
\be
\chi_n \equiv \tr(B^n) = 2\sum_i \cos(n \b_i) \, ,  
\ee
the integration measure is given by 
\be
DB = \prod_m d\chi_m \exp\left( - \sum_{n\; {\rm odd}} \frac{1}{2n} \chi_n^2  
- \sum_{n\;{\rm even}} \frac{1}{2n} (\chi_n +1)^2\right) \,.
\label{sp2nf}
\ee

\paragraph{Result} 
In terms of the shorthand notation, $f_n \equiv f(x^n,y^n)$, 
the index to be computed is 
\be
\CI = \int DA DB \exp\left( \sum_{n=1}^{\infty} \frac{1}{n} f_n \rho_n \chi_n 
\right) \,,
\ee
where the integration measures are given in (\ref{so2nf}) and (\ref{sp2nf}). 
Performing the Gaussian integral by 
diagonalization ($\rho^{\pm}_n = \rho_n \pm \chi_n$), we obtain 
\be
\CI = \prod_{n=1}^{\infty} \frac{1}{\sqrt{1-f_n^2}} 
\prod_{k=1}^{\infty} \exp\left(-\frac{f_{2k}}{2k(1+f_{2k})} \right) \,.
\ee 
The overall normalization is fixed by requiring that $I(x=y=0)=1$.
Using the relations,
\be
\frac{1}{1-f_n^2} = \frac{(1+x^n)^2}{(1-(xy)^n)(1-(x/y)^n)}\,, 
\;\;\;
\frac{f_{2k}}{1+f_{2k}} = \frac{(xy)^k+(x/y)^k}{(1+(xy)^k)(1+(x/y)^k)} \,,
\ee
we can exponentiate the $\prod (1-f_n^2)^{-1/2}$ factor and rewrite the result as
\be
\CI &=& \exp \left( \sum_{k=1}^{\infty} \frac{1}{k} I_{\rm sp} (x^k,y^k) \right) \,,
\nn \\
I_{\rm sp} &=& \frac{x}{1-x^2} + \half \left[\frac{xy}{1-xy}+\frac{x/y}{1-x/y}-\frac{xy+x/y}{(1+xy)(1+x/y)} \right]
\nn \\
&=& \frac{1}{1-x^2} \left[ (1-x/y)\frac{(xy)^2}{1-(xy)^2} 
+(1-xy)\frac{(x/y)^2}{1-(x/y)^2} +x +x^2\right] \,.
\label{ISFT}
\ee

\subsection{Gravity}

The gravity computation can be done by taking the states of $AdS_4 \times S^7$ and keeping the states invariant under the orbifold action. 
The spectrum of $AdS_4 \times S^7$ was originally obtained in \cite{Gunaydin:1985tc,Biran:1983iy} and  recently discussed in the context of the superconformal index in \cite{Bhattacharya:2008zy}.
The $\IZ_k$ orbifolding was studied in \cite{Bhattacharya:2008bja}. 
In the basis where the supercharges transform vectorially under the $SO(8)$, 
the $\IZ_k$ action is a rotation by $4\pi/k$ along the $SO(2)$ part 
of the decomposition $SO(6)\times SO(2) \subset SO(8)$. 
If we denote the generator of the $SO(2)$ by $J_3$, then
the invariant states should satisfy
\be
\label{zk-orbi}
\exp\left(\frac{4\pi i}{k} J_3 \right) | \psi \rangle = 0 \,,
\ee
which means $J_3 | \psi \rangle = 0$ for large $k$. 

The $\ID_k$ group is a subgroup of the $SO(3)$ 
in the decomposition $SO(5)\times SO(3) \subset SO(8)$. 
If we denote the $SO(3)$ generators by $J_{1,2,3}$, 
the generators of $\ID_k$ are given by 
\be
\exp\left(\frac{2\pi i}{k} J_3 \right), \;\;\;
\exp\left( \pi i J_2 \right) \,.
\ee
For large $k$, the first generator again requires that 
$J_3| \psi \rangle = 0$. In the standard $| \ell,m\rangle$ notation, 
all $|\ell \in \IZ , m=0 \rangle$ states satisfy this condition. 
They are also eigenstates of the second generator 
with eigenvalues $(-1)^\ell$. Therefore, the fully invariant states are 
$|\ell \in 2 \IZ , m=0 \rangle$. 

In summary, 
to compute the index over single gravitons, we need to 
decompose the $SO(8)$ graviton spectrum into irreducible representations (irreps) of $SO(5)\times SO(3)$ 
and keep only those $SO(5)$ representations tensored with $|\ell\in 2\IZ,m=0\rangle$ 
states of $SO(3)$. 

In \cite{Bhattacharya:2008bja}, 
the projection from $SO(8)$ to $SO(6)$ was performed 
in two equivalent but slightly different methods.
In the ``indirect'' method, one begins with the index of the unorbifolded  
theory computed in \cite{Bhattacharya:2008zy} 
and projects out the $J_3$-non-invariant states by a contour integral. 
In the ``direct'' method, one takes the graviton spectrum 
of the unorbifolded theory, decomposes them under $SO(8)\rightarrow SO(6)\times SO(2)$ and sums over the $J_3$-invariant subspace of 
the spectrum. 

In the case at hand, the indirect method does not seem to be available. 
On the other hand, the direct method can be implemented without much difficulty. 
We only have to follow the reduction from Table 1 of Ref.~\cite{Bhattacharya:2008zy} 
to Table 3 of Ref.~\cite{Bhattacharya:2008bja}, while taking into account 
the extra condition discussed above. 

\begin{table}
\begin{center}
\begin{tabular}{|c|c|c|c|}
	\hline         
range of $n$ & $\epsilon_0$ & $j$ & $SO(8)$ \  \\
	\hline
$n\geq1$   &$\frac{n}{2}$  &$0$          &($\frac{n}{2},\frac{n}{2},\frac{n}{2},\frac{-n}{2}$)                 \\
$n\geq1$   &$\frac{n+1}{2}$&$\half$      &($\frac{n}{2},\frac{n}{2},\frac{n}{2},\frac{-(n-2)}{2}$)            \\
$n\geq2$   &$\frac{n+2}{2}$&$1$          &($\frac{n}{2},\frac{n}{2},\frac{(n-2)}{2},\frac{-(n-2)}{2}$)         \\
$n\geq2$   &$\frac{n+3}{2}$&$\frac{3}{2}$&($\frac{n}{2},\frac{(n-2)}{2},\frac{(n-2)}{2},\frac{-(n-2)}{2}$)     \\
	\hline
\end{tabular}
\end{center}
\caption{The $\D=0$ subset of the super-graviton spectrum in AdS$_4\times S^7$.}
\label{tb:gravspec}
\end{table}

We reproduce the $\D=0$ subset of Table 1 of Ref.~\cite{Bhattacharya:2008zy} in our Table \ref{tb:gravspec}. Let us first focus on the simplest tower on top of the table. The irreps with half-integer $h_i$ are clearly irrelevant. 
We need to decompose the $(n,n,n,-n)$ reps of $SO(8)$ 
into irreps of $SO(5)\times SO(3)$ and keep only the 
$SO(5)$ states tensored with $|\ell\in 2\IZ,m=0\rangle$ . 
Using the character formulas in appendix \ref{char-so}, we can show that 
\be
\label{chpr}
(n,n,n,-n)_{SO(8)} = \bigoplus_{\ell=0}^{n} (n,\ell)_{SO(5)} \otimes (\ell)_{SO(3)} 
\,.
\ee
So, after the projection, the index receives contributions from 
$(n,2k)$ reps of $SO(5)$ for each integer $n\ge 1$ and $0\le k \le [n/2]$. 
We can decompose the other three towers in Table \ref{tb:gravspec} in a similar manner. The result is as follows.
\bn

\item 
$(n,n,n,-n+1)$ decomposes into four ``families'', 
three of which have $\D=0$:
\[ 
{\rm (a)}\;\;\;
\bigoplus_{\ell=0}^{n-1} (n,\ell+1) \otimes (\ell), 
\;\;\;\;\; 
{\rm (b)}\;\;\;
\bigoplus_{\ell=1}^{n} (n,\ell) \otimes (\ell),
\;\;\;\;\;
{\rm (c)}\;\;\;
\bigoplus_{\ell=1}^{n} (n,\ell-1) \otimes (\ell) .
\]

\item 
$(n,n,n-1,-n+1)$ decomposes into six families, 
three of which have $\D=0$:
\[ 
{\rm (a)}\;\;\;
\bigoplus_{\ell=0}^{n-1} (n,\ell+1) \otimes (\ell), 
\;\;\;\;\; 
{\rm (b)}\;\;\;
\bigoplus_{\ell=1}^{n-1} (n,\ell) \otimes (\ell),
\;\;\;\;\;
{\rm (c)}\;\;\;
\bigoplus_{\ell=1}^{n} (n,\ell-1) \otimes (\ell) .
\]

\item 
$(n,n-1,n-1,-n+1)$ decomposes into four families, 
one of which has $\D=0$:
\[ 
\bigoplus_{\ell=0}^{n-1} (n,\ell) \otimes (\ell).
\]

\en

\noindent
Now, it is easy to sum over all $SO(5)$ reps with $\ell=$(even). 
Introduce the notation, 
\be
Q_1 &\equiv& \sum_{k=0}^{\infty} \chi_{SO(3)}^{(2k)}(y) \sum_{m=0}^{\infty} x^{m+2k} \; = \; \frac{1}{(1-x)(1-y)}\left[\frac{1}{1-x^2/y^2}-\frac{y}{1-x^2y^2}\right] \,, \\
Q_2 &\equiv& \sum_{k=0}^{\infty} \chi_{SO(3)}^{(2k+1)}(y) \sum_{m=0}^{\infty} x^{m+2k}\; = \; \frac{1}{(1-x)(1-y)}\left[\frac{y^{-1}}{1-x^2/y^2}-\frac{y^2}{1-x^2y^2}\right] \,.
\ee
Then, the partial sums
\be
S_j = \sum x^{\e_0+j} \chi_{SO(3)}^{(h)}(y) \,,
\ee
can be written as 
\be
& S_0 = \displaystyle{Q_1-1}\,,& 
\nn \\
S_{1/2}^{(a)} = x^2 Q_2\,, \;\;\;\;\;
&
S_{1/2}^{(b)} = \displaystyle{x \left(Q_1-\frac{1}{1-x} \right)}\,, 
& \;\;\;\;\; 
S_{1/2}^{(c)} = x^3 Q_2\,, 
\nn \\
S_{1}^{(a)} = x^3 Q_2\,, \;\;\;\;\;
&
S_{1}^{(b)} = \displaystyle{ x^3 \left(Q_1-\frac{1}{1-x} \right)}\,, 
& \;\;\;\;\; 
S_{1}^{(c)} = x^4 Q_2\,, 
\nn \\
& S_{3/2} = \displaystyle{ x^4 Q_1} \,.& 
\ee
To sum up, 
the index evaluated over all single gravitons is 
\be
I_{\rm sp} &=& \frac{1}{1-x^2}\sum_j (-1)^{2j} S_j \,. \nn \\
&=& \frac{1}{1-x^2} \left[ (1-x/y)\frac{(xy)^2}{1-(xy)^2} 
+(1-xy)\frac{(x/y)^2}{1-(x/y)^2} +x +x^2\right] \,,
\label{ISGR}
\ee
in perfect agreement with the field theory result (\ref{ISFT}).

\section{Indices for $\cN=4$ theories}

\subsection{$U$-type}

\paragraph{Field Theory}

Since the $R$-symmetry group is $SO(4)$, we define the index to be
\be
\CI(x, y) = \Tr \left[ (-1)^F x^{\epsilon_0 + j} y^{h_2} \right]\,.
\ee
where $h_2$ is the second Cartan charge of $SO(4)$. 
We can read off the letters contributing to the single-letter partition function $f$ from the matter content in Fig.~\ref{quiver}(a). 
The result is summarized in the following table: 
\begin{center}
\begin{tabular}{|c|c|c|c|c|}
\hline
type & operators & $\e_0$ & $j$ & $SO(4)$ \\
\hline
bosons in hyper & $q_+$, $\bar{q}_+$ & $\thalf $ & $0$ &  $(\thalf,\thalf)$ \\
fermions in hyper & $\psi_+$, $\bar{\psi}_+$ & $1$ & $\thalf$ &  $(\thalf, -\thalf)$ \\
\hline
bosons in twisted hyper & $\tilde{q}_+$, $\bar{\tilde{q}}_+$ & $\thalf$ & $0$ &  $(\thalf, -\thalf)$ \\
fermions in twisted hyper& $\tilde{\psi}_+$,$\bar{\tilde{\psi}}_+$ & $1$ & $\thalf$ &  $(\thalf,\thalf)$ \\
\hline 
derivatives & $\partial$ & $1$ & $1$ & $(0,0)$ \\
\hline 
\end{tabular}
\end{center}
Then, the single letter partition function is given by
\be
{\rm hyper} &:& f(x, y) = \frac{x^{1/2} y^{1/2}}{1-x^2} - \frac{x^{3/2} y^{-1/2}}{1-x^2} = \frac{\sqrt{xy}}{1-x^2} \left(1 - \frac{x}{y} \right) \,,
\\
{\rm twisted}\;\; {\rm hyper} &:& \tilde{f}(x, y) = \frac{x^{1/2} y^{-1/2}}{1-x^2} - \frac{x^{3/2} y^{1/2}}{1-x^2} = \frac{\sqrt{x/y}}{1-x^2} \left(1 - xy \right) \,.
\ee
To calculate the index, we have to evaluate the following matrix integral.
\be
\CI^U_{m} &=& \int \prod_{i=1}^{2m} DU_i \exp \left( \sum_{n=1}^{\infty} \sum_{i=1}^m \frac{1}{n} F_{n,i} \right) \,,
\nn \\
F_{n,i} &=& f_n \left[ \Tr ( U_{2i-1}^n ) \Tr( U_{2i}^{-n} ) + \Tr ( U_{2i-1}^{-n} ) \Tr( U_{2i}^{n} ) \right] 
\nn \\
&& +  \tilde{f}_n \left[ \Tr ( U_{2i}^n ) \Tr( U_{2i+1}^{-n}) +  \Tr ( U_{2i}^{-n} ) \Tr( U_{2i+1}^{n}) \right] \,.
\ee
Here, we used the shorthand notation $f_n = f(x^n, y^n)$ and $\tilde{f}_n = \tilde{f}(x^n, y^n)$, and $U_{2m+1}$ is identified with $U_1$.
By changing the variables to $\rho_{i, n} = \frac{1}{N} \Tr(U_{2i-1}^n)$ and $\chi_{i, n} = \frac{1}{N} \Tr(U_{2i}^n)$ and using 
the measure of the matrix integral, we can rewrite the integral as 
\be
\CI^U_{m} &=& \int \prod_{i=1}^m \prod_{n=1}^{\infty} d\rho_{i, n} d\chi_{i, n}  
 \exp \left( -N^2 \sum_{i, n} \frac{1}{n} \left[  |\rho_{i, n}|^2 +  |\chi_{i, n}|^2 - G_{n,i} \right] \right) \,, 
\nn \\
G_{n,i} &=& f_n \rho_{i, n} \chi_{i, n} 
   + f_n \rho_{i, -n} \chi_{i, n} 
  + \tilde{f}_n \chi_{i, n} \rho_{i+1, -n} 
  + \tilde{f}_n \chi_{i, -n} \rho_{i+1, n} \,.
\ee
As in the $\CN=6$ case, it is convenient to write the integral in a matrix form,  
\be
\CI^U_{m} = \int \prod_{i=1}^m \prod_{n=1}^{\infty} d\rho_{i, n} d\chi_{i, n} \exp \left( -\frac{N^2}{2} \sum_{n=1}^{\infty} \frac{1}{n} \left( C_{m,n}^T M_{m,n} C_{m,n}  \right) \right) \,.
\ee
Here, $C_{m,n}^T = (\rho_{1, n} ~ \chi_{1, n} ~ \rho_{1, -n} ~ \chi_{1, -n} ~\cdots~ \rho_{m,-n}~\chi_{m, -n})$, and $M_{m,n}$ is a positive definite $4m \times 4m$ matrix which coincides with (\ref{m144}) when $m=1$. 
Performing the integral, we find
\be
\CI^U_{m} = \prod_{n=1}^{\infty} \frac{(1-x^{2n})^{2m}}{(1-x^n y^n)^m(1-x^n/y^n)^m(1-x^{n m})^2} \,.
\ee
The corresponding single-particle index is 
\be
\label{am-ftsp}
I^U_{{\rm sp},{m}} (x,y) = m\left( \frac{1}{1-xy} + \frac{1}{1-x/y} -\frac{2}{1-x^2}\right) + \frac{2 x^{m}}{1-x^{m}} \,.
\ee

\paragraph{Gravity}
The gravity computation in the untwisted sector turns out to be almost trivial. 
The orbifolding action due to the Chern-Simons level (in the $k\goto\infty$ limit) makes the same effect as in the process of going from $\cN=8$ to $\cN=6$. Thus, we can begin with the $\cN=6$ single-particle index 
\be
I_{\rm sp}^{\cN=6}(x,y_1,y_2) = \frac{x}{y_1-x}+\frac{1}{1-x y_1}+\frac{x}{y_2-x}+\frac{1}{1-x y_2}- \frac{2}{1-x^2} \,.
\ee
The other $\IZ_{m}$ orbifolding acts only on $y_2$. So, we can simply take
\be
I_{{\rm sp}, m}^U(x,y) \;=\; \frac{1}{m}\sum_{j=1}^m I_{\rm sp}^{\cN=6}(x,y,\w_m^j) 
\;=\;
\frac{1}{1-xy} + \frac{1}{1-x/y} -\frac{2}{1-x^2} + \frac{2 x^{m}}{1-x^{m}} \,.
\label{am-gravsp}
\ee
Comparing (\ref{am-gravsp}) and (\ref{am-ftsp}), 
we find a mismatch 
\be
\D I^U_{{\rm sp},m} &=& (m-1) \left( \frac{1}{1-xy} + \frac{1}{1-x/y} -\frac{2}{1-x^2}\right) 
\nn \\
&=&
(m-1)
\frac{1}{1-x^2}\left((1-x/y)\frac{xy}{1-xy} + (1-xy)\frac{x/y}{1-x/y}\right)
\,.
\label{mis-u}
\ee
We will now argue that the twisted sector contributions can account for 
the mismatch and lead to perfect agreement of the index between field theory and gravity.  

\paragraph{Twisted sector $-$ field theory}

Suppose we have operators $O(i)$ $(i=1,\cdots,m)$ which form 
a regular representation of $\IZ_m$. Then, the $(m-1)$ linearly independent 
operators $O(i+1)-O(i)$ not invariant under $\IZ_m$ must belong 
to the twisted sector. 

The bosonic single-trace operators that contribute to the index are given by
\be
 O_n^B(i) \; = \;
 \Tr \left(q^i_+ \bar{q}^i_+\right)^{n} \,,
\;\;\;\;\;
 \widetilde{O}^B_n(i) \; = \; 
 \Tr \left(\tilde{q}^i_+ \bar{\tilde{q}}^i_+ \right)^n \,,
 \label{twi-boson}
\ee
where $q^i$ are hypers in $(N,\bar{N})$ of $U(N)_{2i-1}$ and $U(N)_{2i}$, 
and $\tilde{q}^i$s are twisted hypers in $(N,\bar{N})$ of $U(N)_{2i} \times U(N)_{2i+1}$. The bars denote hermitian conjugation. 
As explained earlier, the subscripts $(\pm)$ denote the doublet indices under the $SU(2)_L\times SU(2)_R$ $R$-symmetry. 

The supersymmetry transformation rule (\ref{n4susy}) shows that 
these operators are annihilated by $Q_{++}$ and $S_{--} = (Q_{++})^\dagger$, so they contribute to the index. 
Their quantum numbers are $(\e_0,j,h_1,h_2)= (n, 0, n, \pm n)$, 
so $\D = \e_0-j-h_1=0$ as expected.

To obtain the fermionic operators, we can take the super-descendants of the bosonic operators by acting with supercharges commuting with $Q_{++}$. We find
\be
 O^F_n(i) &=& \left[ Q_{-+} , O^B_n(i) \right] \;=\;
 \Tr \left[ (\psi^i_+ \bar{q}^i_+ + q^i_+ \bar{\psi}^i_+ )(q^i_+\bar{q}_+^i)^{n-1} \right] \,,
 \nn \\
 \widetilde{O}^F_n (i) &=& \left[ Q_{+-} , \widetilde{O}^B_n(i) \right] =
 \Tr \left[ (\tilde{\psi}^i_+ \bar{\tilde{q}}^i_+ + \tilde{q}^i_+ \bar{\tilde{\psi}}^i_+)(\tilde{q}^i_+ \bar{\tilde{q}}^i_+ )^{n-1}\right]  \,,
 \label{twi-ferm}
\ee 
whose quantum numbers are $(\e_0,j,h_1,h_2)=(n+1/2,1/2,n,\pm (n-1))$ such that $\D=0$. The superconformal algebra,  
\be
\{ Q_{++} , Q_{\pm \mp} \} = 0, 
\;\;\; 
\{ S_{--}, Q_{-+} \} \propto J_{--}, 
\;\;\; 
\{ S_{--}, Q_{+-} \} \propto \tilde{J}_{--}, 
\ee
where $J$ and $\tilde{J}$ are generators of $SU(2)_L\times SU(2)_R$, 
shows that these fermionic operators are also annihilated 
by $Q_{++}$ and $S_{--}$ and contribute to the index. 
Taking into account the bosonic descendants (derivatives), we obtain the index summed over the $\IZ_{m}$ non-invariant bosonic and fermionic operators, 
\be 
I^U_{{\rm sp},m} {\rm (twisted)} = (m-1) \times \frac{1}{1-x^2}\left[(1-x/y)\frac{xy}{1-xy} + (1-xy)\frac{x/y}{1-x/y}\right] \,,
\label{twi-u}
\ee 
which agrees precisely with the mismatch (\ref{mis-u}). 

It is rather remarkable that (\ref{twi-boson}) and (\ref{twi-ferm}) exhaust all twisted sector contributions to the index, as 
there are many more operators which apparently satisfy $\D=0$. 
However, all such operators connecting three or more nodes 
of the quiver can be shown to be $(Q_{++})$-exact 
by using the supersymmetry transformation rule (\ref{n4susy}).
For instance, repeated use of (\ref{n4susy}) shows that 
\be
&&O(i) = \Tr( q^i_+ \bar{q}^i_+ q^{i}_+ \tilde{q}^{i}_+ \bar{\tilde{q}}^i_+ \bar{q}^i_+ ) \,,
\nn \\
&&O(i+1) - O(i) = [ Q_{++}, \Tr( q^{i+1}_+ \bar{q}^{i+1}_+ q^{i+1}_+ \bar{\psi}^{i+1}_+  + 
q^{i+1}_+ \bar{q}^{i+1}_+ \bar{\tilde{\psi}}^{i}_+ \tilde{q}^i + \bar{\tilde{\psi}}_+^i \bar{q}_+^i q_+^i \tilde{q}_+^i) 
] \,.
\ee
In the $\CN=2$ superfield language adopted in \cite{Benna,Tera}, 
this is a consequence of the F-term equivalence relations. 

\paragraph{Twisted sector $-$ gravity}

The field theory computation above implies that 
there should be corresponding twisted sector states on the gravity side. 
In fact, the orbifold has two copies of $S^2 \subset \IC \IP^3 \subset S^7$ as the fixed loci, one at $(z_1, z_2,0,0)$ and the other at $(0,0,z_3,z_4)$, as the circle fiber of the Hopf fibration of $S^3$ is removed by the $\IZ_{mk}$ orbifold action in the large $k$ limit. 
 Therefore, in type IIA description, there should exist chiral primary states in the twisted sector localized at these fixed loci.  
Note that the splitting of twisted sector states into 
hypers and twisted hypers as in (\ref{twi-boson}) and (\ref{twi-ferm}) 
matches nicely with two disjoint fixed loci of the orbifold geometry.

To understand the nature of the twisted sector states, let us examine 
the geometry near the fixed loci. We begin with writing the metric of 
$S^7$ as 
\be \label{s7metric}
ds^2 &=& d\a^2 +\sin^2\a d\O_1^2 + \cos^2\a d\O_2^2  
\nn \\
&=& d\a^2 + \textstyle{\frac{1}{4}}\sin^2\a \left[ d\th_1^2+ \sin^2\th_1 d\phi_1^2 + (d\psi_1 +\cos\th_1 d\phi_1)^2) \right] 
\nn \\
&& \;\;\;\;\;\;\;\;
+ \textstyle{\frac{1}{4}}\cos^2\a \left[ d\th_2^2+ \sin^2\th_2 d\phi_2^2 + (d\psi_2 +\cos\th_2 d\phi_2)^2) \right] \,.
\ee
The orbifold action can be taken to be
\be
(\psi_1, \psi_2) \sim (\psi_1 + 2\pi/m, \psi_2) \sim (\psi_1 + 2\pi/mk, \psi_2+ 2\pi/mk) \,.
\ee
Near $\a = 0$, it is convenient to take the ``fundamental domain'' of 
the $(\psi_1, \psi_2)$ torus as follows:
\be
\psi_1 = \psi + \b , \;\;\; \psi_2 = \psi , \;\;\; 
0 < \psi < 2\pi/mk, \;\;\; 0 < \beta < 2\pi /m . 
\ee 
Then the metric can be rewritten as
\be
ds^2  
&=& d\a^2 + \textstyle{\frac{1}{4}}\sin^2\a \left( d\th_1^2+ \sin^2\th_1 d\phi_1^2 
\right) 
+ \textstyle{\frac{1}{4}}\cos^2\a \left( d\th_2^2+ \sin^2\th_2 d\phi_2^2  \right) 
\nn \\
&&+\textstyle{\frac{1}{4}} \cos^2\a\sin^2\a(d\b+\cos\th_1d\phi_1 - \cos\th_2 d\phi_2)^2 
\nn \\
&& + \textstyle{\frac{1}{4}}\left[d\psi+  \sin^2\a (d\b+ \cos\th_1d\phi_1) + \cos^2\a \cos\th_2 d\phi_2\right]^2 
\nn \\ 
&\approx& d\a^2+ \textstyle{\frac{1}{4}} \a^2 \left[ d\th_1^2+ \sin^2\th_1 d\phi_1^2 +(d\b+\cos\th_1d\phi_1 - \cos\th_2 d\phi_2)^2 \right]
\nn \\
&&
+ \textstyle{\frac{1}{4}} \left( d\th_2^2+ \sin^2\th_2 d\phi_2^2  \right) 
+ \textstyle{\frac{1}{4}}\left[d\psi+  \cdots \right]^2 \,.
\label{orbi-met}
\ee
The angle $\psi$ is the coordinate of the 11-th circle. So, up to an overall warping, 
the IIA metric near the singularity looks like a non-trivial fibration of $A_{m-1}$ over $S^2$. 

A direct string theoretic analysis of the twisted sector states would be a formidable task. Instead, we could follow the analysis of \cite{Gukov:1998kk} 
where blow-up of the orbifold was used to obtain the spectrum 
of the twisted sector states. There are $(m-1)$ normalizable harmonic two-forms 
$\w_i$ in the blown-up $A_{m-1}$ singularity. 
As in \cite{Gukov:1998kk}, the candidate for chiral primary states 
in the twisted sector comes from the harmonic decomposition of the NSNS $B$-field into 
$(m-1)$ scalars by
\be
B = \sum_{i=1}^{m-1} \phi_i \,\w_i \,.
\ee 
The spherical harmonics on either $S^2$ 
have integer ``spin''-$n$ under $SU(2)_L$ or $SU(2)_R$ 
factors of the $SO(4)$ $R$-symmetry. 
In particular, the highest weight states have the $SO(4)$ 
Cartan charges $(h_1,h_2)= (n,\pm n)$. 
Therefore, in order to match the quantum numbers 
$(\e_0,j,h_1,h_2)=(n,0,n,\pm n)$ of the twisted sector states 
found in the field theory, 
it would be sufficient to show that these states have $\e_0=n$, 
which is equivalent to $({\rm mass})^2 = n(n-3)$. 
The Laplacian operator contributes $n(n+1)$ to the mass squared of the scalar fields. As noted in \cite{Gukov:1998kk}, the interaction of the $B$-field with background RR 3-form and 1-form fields is likely to produce 
a shift in mass squared. Having $\e_0=n$ amounts to a mass shift 
$\d({\rm mass})^2 = -4 n$. 

The computation of the mass shift would take two steps. 
First, we need to blow up the metric (\ref{orbi-met}) while maintaining 
the Einstein condition. Second, we solve the equation,  
\be
d*dC_3 = \half dC_3 \wedge dC_3\,,
\ee
of eleven-dimensional supergravity with the ansatz
\be
C_3 = C_3({\rm background}) +\phi_i \, \w_i \wedge (d\psi + \cdots) \,.
\ee
We leave the detailed computation of the mass shift to a future work.

\subsection{$OSp$-type}

\paragraph{Field Theory} We have $m$ copies of alternating $O(2N)$ and $Sp(2N)$ gauge groups, and the quiver diagram is given in figure \ref{quiver}(b). Since the matter content is all the same as in 
the $U$-type theory except that they are in the real representations of the gauge group, the single letter partition function is exactly the same as that of the $U$-type theory. 
\be
{\rm hyper} &:& f(x, y) = \frac{x^{1/2} y^{1/2}}{1-x^2} - \frac{x^{3/2} y^{-1/2}}{1-x^2} = \frac{\sqrt{xy}}{1-x^2} \left(1 - \frac{x}{y} \right) \,,
\\
{\rm twisted}\;\; {\rm hyper} &:& \tilde{f}(x, y) = \frac{x^{1/2} y^{-1/2}}{1-x^2} - \frac{x^{3/2} y^{1/2}}{1-x^2} = \frac{\sqrt{x/y}}{1-x^2} \left(1 - xy \right) \,.
\ee
Plugging in the single letter partition functions to the matrix-integral formula for the index, we obtain,
\be
\CI^{OSp}_{m} &=& \int \prod_{i=1}^{m} DA_i DB_i \exp \left( \sum_{n=1}^{\infty} \sum_{i=1}^m \frac{1}{n} F_{n,i} \right) \,,
\nn \\
F_{n,i} &=& f_n  \Tr ( A_{i}^n ) \Tr( B_{i}^{n} ) + \tilde{f_n} \Tr ( B_{i}^n ) \Tr( A_{i+1}^{n}) \,,
\ee
where $f_n=f(x^n, y^n)$ and $A$ and $B$ are elements of orthogonal and symplectic gauge group respectively.
Since we consider circular quivers, we identify $A_{m+1}\sim A_{1}$ and $B_{m+1}\sim B_1$. We can repeat the same matrix integral for $\CN=5$ $OSp$ theory except that there are $m$ copies of gauge group and matters. 
The result is
\be
\CI^{OSp}_m = \prod_{n=1} \frac{(1-x^{2n})^m}{(1-(xy)^n)^{m/2}(1-(x/y)^n)^{m/2}(1-x^{nm})} \prod_{k=1} \exp{\left(-\frac{f_{2k}}{2k(1+f_{2k})}\right)} .
\ee
The corresponding single-particle index is
\be
\label{omk-ftsp}
I_{{\rm sp},m}^{OSp}= \frac{m}{2}\left( \frac{1}{1-xy} + \frac{1}{1-x/y} -\frac{2}{1-x^2}-\frac{xy+x/y}{(1+xy)(1+x/y)}\right) + \frac{x^{m}}{1-x^{m}} \,.
\ee
For $m=1$, it agrees with the $\CN=5$ result as expected.

\paragraph{Gravity} Again, we can work out the decomposition of 
representations, etc., but there is a shortcut. 
Consider the broken $SO(4)= SU(2)\times SU(2) \subset SO(8)$ and denote 
its representations by $(j_1,j_2)$. 
Taking the $\ID_k$ orbifold action into account, but 
leaving the $\IZ_m$ orbifolding aside, we can write down 
the ``preliminary version'' of the index as follows:
\be
\label{omk-grav1}
I_{{\rm sp}, m}^{OSp-{\rm pre}} = \int \frac{d\th}{2\pi} \tr \left[ \frac{x^{\e_0+j}}{1-x^2} y_1^{h_2} y_2^{h_3}  \left(\frac{1+e^{\pi i \left\{J_2^{(1)}+J_2^{(2)}\right\}} }{2} \right) e^{i \th \left\{ J_3^{(1)}+ J_3^{(2)}\right\}}  \right] \,.
\ee
As before, 
the $\IZ_{k}$ orbifolding picks out only the states with $m_1+m_2=0$. 
The other generator of $\ID_k$ acts on these states as 
\be
\label{dkeff}
e^{\pi i \left[J_2^{(1)}+J_2^{(2)}\right]} |j_1,j_2; +m,-m \rangle = (-1)^{j_1+j_2} 
|j_1,j_2; -m,+m \rangle \,.
\ee
The first half of (\ref{omk-grav1})  
which does not involve $J_2^{(1)}+J_2^{(2)}$ is nothing but 
one half of the $\cN=6$ index. 
The other half (call it $\D I$) picks up contributions 
only from the states with $m_1 = m_2 = 0$ because of (\ref{dkeff}).
Since $h_3 = m_1 - m_2$ in (\ref{omk-grav1}), $\D I$ is independent of $y_2$. 

Before the $\IZ_m$ orbifolding, the index must coincide with 
the $\cN=5$ result.\footnote{
We could reverse the logic, compute $\D I$ directly 
and give an alternative derivation of the $\cN=5$ index, 
but we will not pursue it here.}
Therefore, 
\be
\D I(x,y) = I_{\rm sp}^{\cN=5}(x,y) - \thalf I_{\rm sp}^{\cN=6}(x,y,y_2=1) \,.
\ee
The $\IZ_m$ orbifolding acts only on $y_2$, so it makes no effect 
on $\D I$. 
\be
\label{omk-grav2}
I_{{\rm sp}, m}^{OSp}(x,y) &=& \frac{1}{m} \sum_{j=1}^m I_{{\rm sp}, 1}^{OSp-{\rm pre}}(x,y,\w_m^j) \;\;\;\;\; (\w_m \equiv e^{2\pi i/m})
\nn \\
&=& \frac{1}{2m} \sum_{j=1}^m I_{\rm sp}^{\cN=6}(x,y,\w_m^j) +\D I(x,y)  
\\
&=& \frac{1}{2}\left( \frac{1}{1-xy} + \frac{1}{1-x/y} -\frac{2}{1-x^2}-\frac{xy+x/y}{(1+xy)(1+x/y)}\right) + \frac{x^{m}}{1-x^{m}} \,.
\nn
\ee
Comparing (\ref{omk-ftsp}) and (\ref{omk-grav2}), 
we find a mismatch
\be
\Delta I_{{\rm sp}, m}^{OSp}(x,y) &=& \frac{m-1}{2} \left( \frac{1}{1-xy} + \frac{1}{1-x/y} -\frac{2}{1-x^2}-\frac{xy+x/y}{(1+xy)(1+x/y)}\right)
\nn \\ 
&=& 
(m-1) \times {1-x^2} \left[ (1-x/y)\frac{(xy)^2}{1-(xy)^2} 
+(1-xy)\frac{(x/y)^2}{1-(x/y)^2} \right] \,.
\label{mis-osp}
\ee 

\paragraph{Twisted sector}

We can repeat the same analysis as in the previous case with slight modifications. The bosons contributing to the index are given by 
\be
 O_n^B(i) \; = \;
 \Tr \left(q^i_+ \bar{q}^i_+\right)^{2n} \,,
\;\;\;\;\;
 \widetilde{O}^B_n(i) \; = \; 
 \Tr \left(\tilde{q}^i_+ \bar{\tilde{q}}^i_+ \right)^{2n}  \,.
 \label{twi-boson2}
\ee
Recall that $q$ and $\bar{q}$ are related by $\bar{q}^{a\dot{a}} = \delta^{ab} \w^{\dot{a}\dot{b}}q_{b\dot{b}}$. We have $q^{4n}$ rather than $q^{2n}$ because the trace vanishes due to the anti-symmetric tensor $\w^{\dot{a}\dot{b}}$ of $Sp(2N)$. 

The fermionic operators corresponding to the index 
are again obtained by taking super-descendants of the bosonic operators: 
\be
 O^F_n(i) &=& \left[ Q_{-+} , O^B_n(i) \right] \;=\;
 \Tr \left[ (\psi^i_+ \bar{q}^i_+ + q^i_+ \bar{\psi}^i_+ )(q^i_+\bar{q}_+^i)^{2n-1} \right] \,,
 \nn \\
 \widetilde{O}^F_n (i) &=& \left[ Q_{+-} , \widetilde{O}^B_n(i) \right] =
 \Tr \left[ (\tilde{\psi}^i_+ \bar{\tilde{q}}^i_+ + \tilde{q}^i_+ \bar{\tilde{\psi}}^i_+)(\tilde{q}^i_+ \bar{\tilde{q}}^i_+ )^{2n-1}\right] .
\ee 
The bosonic and fermionic operators together contribute to the index by
\be 
(m-1) \times \frac{1}{1-x^2}\left[(1-x/y)\frac{(xy)^2}{1-(xy)^2}+(1-xy)\frac{(x/y)^2}{1-(x/y)^2}
\right] \,,
\ee 
which exactly account for the difference between supergravity approximation and field theory computation (\ref{mis-osp}) . 

On the gravity side, we again have fixed loci at $(z_1,z_2,0,0)$ 
and $(0,0,z_3,z_4)$. 
The $\ID_{k}$ action
$z_i \sim \omega_k z_i$ and $(z_1, z_2, z_3, z_4) \sim (\bar{z_2}, -\bar{z_1}, \bar {z_3}, -\bar{z_4})$ make the fixed loci 
to be $\IR\IP^2=S^2/\IZ_2$. Among the spherical harmonics 
on $S^2$, only those with even $n$ contribute. 
This matches the spectrum of twisted sector states in the field theory.

\subsection{On the dual ABJM proposal}

In \cite{Hanany:2008fj,Franco:2008um}, a $U(N) \times U(N)$ Chern-Simons theory with manifest $\cN=2$ supersymmetry and the superpotential $W = [\phi_1, \phi_2]AB$ was proposed, where $\phi_1, \phi_2$ are adjoints in the first gauge group and $A, B$ are bi-fundamentals. The moduli space of vacua was claimed to be $\IC^2/\IZ_k \times \IC^2$, which suggests that the theory may exhibit $\cN=4$ at the IR fixed point. The existence of non-trivial fixed point for $\cN=2$ Chern-Simons-Matter theories were shown in \cite{Gaiotto:2007qi}. The moduli space seems to imply that the IR fixed point of this theory is dual to M-theory on AdS$_4 \times S^7/\IZ_k$ where the $\IZ_k$ acts on either $\psi_1$ or $\psi_2$ in \eqref{s7metric}.  For $k=1$, the moduli space is $\IC^4$, and the authors of \cite{Hanany:2008fj,Franco:2008um} claim that the theory flow to the same IR fixed point as the ABJM theory for $k=1$ (hence the name ``dual ABJM''). It would be interesting to test whether the fixed point theory actually has $\cN=4$ supersymmetry when $k > 1$. For this purpose, we will compute the index on both sides and make comparison. 

All the field theory computations in previous sections were based 
on the fact that we can take the free theory limit by continuous marginal deformation. The free theory computation cannot be justified 
if the theory undergoes a non-trivial RG flow. 
However, the analysis \cite{Gaiotto:2007qi} of the RG fixed point 
of $\CN=2$ theories with quartic superpotential 
indicates that the coefficient of the superpotential 
at the fixed point is of order $1/k$. So, in the large $k$ limit, 
the theory is weakly coupled throughout the RG-flow 
and the free theory computation may be reliable.  
We will proceed to perform the 
free theory computation and check the consistency of 
the approach a posteriori.

\paragraph{Gravity} 
As usual, in the $\cN=8$ setup before orbifolding, we take the basis of $SO(8)$ such that the supercharge 
$Q$ has highest weight $(1,0,0,0)$ and the scalar field has 
highest weight $(1/2,1/2,1/2,-1/2)$. 
Let's denote the 7-sphere to be 
\be
S^7 = \{ (x_1, \cdots, x_8) | \sum_{i=1}^8 |x_i|^2 = 1 \} .
\ee
Consider the $\IZ_k$ orbifolding action by
\be
\exp\left( \frac{2\pi}{k} (H_{56}+H_{78})\right) \,,
\ee
where $H_{56}, H_{78}$ denotes rotation generators in $x_5-x_6$ and $x_7-x_8$ planes. Clearly, the surviving supercharges are vectors in $x_{1, 2, 3, 4}$ directions.  
Under $SO(8)$, the scalar fields decompose as
\be \label{scalardec}
\phi_1 &:& \pm\thalf(+,+,+,-) \,,
\nn \\
\phi_2 &:& \pm\thalf(+,+,-,+) \,,
\nn \\
A &:& \pm\thalf(+,-,+,+) \,,
\nn \\
B &:& \pm\thalf(+,-,-,-) \,.
\ee
Note that $\phi_{1,2}$ are neutral under the orbifold action, 
while $A$ and $B$ have charge $\pm 1$. 

To incorporate the effect orbifolding in the index, 
we can invoke the contour integral method. 
Take $I_{\rm sp}^{\cN=8}(x,y_1,y_2,y_3)$, set $y_2=y_3=z$ and collect 
the $z$-independent terms by using the contour integral. 
The result is (setting $\sqrt{x} = t$ and $\sqrt{y_1} =u$)
\be
I_{\rm sp}^{\rm gravity} = -1 + \frac{1}{(1-tu)^2}-\frac{2t^4}{(1-t^4)(1-tu)}-\frac{t^2}{t^2-u^2} \,.
\ee

\paragraph{Field Theory}
From \eqref{scalardec}, we can identify the letters contributing to the index as follows. 
\begin{center}
\begin{tabular}{|c|c|c|c|c|}
\hline
type & operators & $\e_0$ & $j$ & $(h_1, h_2)$ \\
\hline
bosons & $\phi_{1, b}$ & $\thalf $ & $0$ &  $(\thalf,\thalf)$ \\
& $\phi_{2, b}$ & $\thalf $ & $0$ &  $(\thalf, \thalf)$ \\
& $A_b$ & $\thalf $ & $0$ &  $(\thalf,-\thalf)$ \\
& $B_b$ & $\thalf $ & $0$ &  $(\thalf,-\thalf)$ \\
\hline 
fermions & $\phi_{1, f}$ & $1 $ & $\thalf$ &  $(\thalf, -\thalf)$ \\
& $\phi_{2, f}$ & $1 $ & $\thalf$ &  $(\thalf, -\thalf)$ \\
& $A_f$ & $1 $ & $\thalf$ &  $(\thalf,\thalf)$ \\
& $B_f$ & $1$ & $\thalf$ &  $(\thalf,\thalf)$ \\
\hline 
derivatives & $\partial$ & $1$ & $1$ & $(0,0)$ \\
\hline 
\end{tabular}
\end{center}
Here, the subscript $b, f$ denotes the bosonic and fermionic components of the corresponding superfield. 
The ``indices for letters'' are given by 
\be
f_{12} = f_{21} = \frac{t/u-t^3u}{1-t^4}, 
\;\;\;
f_{11} = \frac{2(tu-t^3/u)}{1-t^4}, 
\;\;\; 
f_{22} = 0 \,.
\ee 
The matrix integral is straightforward to compute. 
\be
\CI^{\rm FT} &=& \int DU_1 DU_2 \exp\left( \sum_{n=1}^{\infty} \sum_{a,b} \frac{1}{n} f_{ab}(t^n, u^n) \tr(U_a^n)\tr(U_b^{-n}) \right)
\nn \\
&=& \prod_{n=1}^{\infty} \frac{1}{1-f_{11}^{(n)}-f_{12}^{(n)}f_{21}^{(n)}}
\nn \\
&=& \prod_{n=1}^{\infty} \frac{(1-t^{4n})^2}{(1-t^{2n}/u^{2n})(1-2t^nu^n + 2t^{5n}u^n-t^{6n}u^{2n})} \,.
\ee

\paragraph{Comparison}
Let's expand the index by powers of $t$. The field theory index is given by
\be
 \CI^{\rm FT} &=& 1 + 2 u t + (u^{-2} + 6u^2) t^2 + (2u^{-1} + 14 u^3) t^3 + 
 (2 u^{-4} + 4 + 34 u^4) t^4 \nonumber \\ 
 &+& (4u^{-3} + 8 u + 74 u^5)t^5 +(3 u^{-6} + 10 u^{-2} + 15 u^2 + 166 u^6) t^6  + \cdots,
\ee
and the gravity side is given by
\be
\CI^{\rm gravity} &=& \exp\left( \sum_{k=1}^{\infty} \frac{1}{k} I_{\rm sp}^{\rm gravity}(t^n, u^n) \right) \nonumber \\
 &=& 1 + 2 u t + (u^{-2} + 6u^2) t^2 + (2u^{-1} + 14 u^3) t^3 + 
  (2 u^{-4} + 4 + 33 u^4) t^4 \nonumber \\ 
 &&+ (4u^{-3} + 8 u + 70 u^5)t^5 +(3 u^{-6} + 10 u^{-2} + 15 u^2 + 149 u^6) t^6  + \cdots,
\ee
The results match to a large extent, but not completely:
\be
\CI^{\rm FT} - \CI^{\rm gravity} = 
u^4t^4 + 4 u^5t^5+ 17 u^6t^6 + \cdots\,.
\ee
Note that $\CI^{\rm FT} > \CI^{\rm gravity}$ holds order by order
and that the the orbifold has the fixed locus $S^3 \subset S^7/\IZ_k$. 
So, it seems possible for the twisted sector states to account for 
the mismatch. It is not clear, however, why only a small number 
of terms survive in the $k\goto \infty$ limit. 
It would be worthwhile to repeat the computation 
for large but finite $k$, where perturbation theory is applicable, 
in order to fully verify the existence of the $\cN=4$ fixed point 
of the dual ABJM theory. 

%
%
\section{Conclusion}
We have checked that the superconformal indices of $\cN=4, 5$ Chern-Simons theories match with those of their gravitational dual theories. This provides a non-trivial check of the proposed dual geometry. Especially, from the fact that orbifold has fixed locus, and by identifying the contributions from the twisted sectors, we can verify that the superconformal theory is not dual to just supergravity, but full superstring theory. Also, we have computed the index for the dual ABJM model proposed by \cite{Hanany:2008fj}. Our computation suggests that we have to go beyond the supergravity approximation to actually verify the claim. 

In principle, the superconformal index depends both on $N$ and $k$, because they determine the form of possible gauge invariant operators. Our computation was done in the large $k$ and large $N$ limit. In this regime, we can use weakly coupled type IIA superstring description. 
We can regard the 't Hooft coupling $\lambda = N/k $ to be effectively continuous so that the index does not change as we go from weak coupling (free field theory) to very strong coupling (supergravity). But, the index is still discrete for finite $N$, so it remains to be checked whether this index does not `jump' under such deformations. In addition, to properly understand the behavior of the index when the two gauge groups have different ranks, ({\it e.g.} $U(M) \times U(N)$, $M \ne N$) we should look into the case of finite $N$. According to \cite{Aharony:2008gk}, $\cN=6$ $U(M) \times U(N)$ Chern-Simons theory remains superconformal only if $|M-N| \le k$. To check this, we should also investigate the finite $k$ effect. On the gravity side, finite $k$ effect is easily computable for large $N$, where we can use supergravity approximation. But in order to see finite $N$ effect, we should compute $1/N$ corrections. We leave this as a future work. 
\footnote{ 
Matching of the index of finite rank gauge groups in four dimensions 
was previously reported in \cite{Dolan2}.} 

It is possible to compute the superconformal indices for less supersymmetric Chern-Simons theories. But, less supersymmetry means the index has fewer expansion parameters. For example, in the case of $\cN=2, 3$ theories, we only have 1 parameter in the index. And for $\cN=1$ theories, it becomes an  ordinary Witten index. So as we go to less supersymmetic case, the index contains less information. Also, the gravity duals of less supersymmetric theories are more complicated, and the KK spectrum of the compact space is not known. However, it is worth noting that the superconformal index is invariant under any deformation that preserves the special supercharges $Q$ and $S$. In our examples presented here, there is no marginal deformation preserving the same number of supersymmetry besides the Lagrangian itself. But, there are possible deformations which preserve less supersymmetries. In this case, we should be able to identify the corresponding deformation of the dual geometry on the gravity side. One possible example is the $\cN=1$ Chern-Simons theory obtained from deforming the free theory limit of $\cN=6$ ABJM theory \cite{Ooguri:2008dk}. The dual geometry of the theory is conjectured to be the orbifolded squashed seven-sphere. Since the free theory limit of the both theory agrees, it seems possible to construct the index which includes the global symmetry along with superconformal algebra. 
Another example is the three dimensional version of the beta-deformation 
preserving $\CN=2$ supersymmetry. 
The gravity dual of the beta-deformation was studied earlier in 
\cite{Lunin, Ahn, Gaunt}. It would be interesting to study the deformation 
in the corresponding Chern-Simons theories.


%
%

\acknowledgments
We are grateful to Seok Kim, Sungjay Lee, Hirosi Ooguri, Chang-Soon Park and Masahito Yamazaki for helpful discussions. 
The work of S.L. is supported in part by the KOSEF Grant
R01-2006-000-10965-0 and the Korea Research Foundation Grant
KRF-2007-331-C00073. The work of J.S. is supported in 
part by  DOE 
grant DE-FG02-92ER40701 and Samsung Scholarship.

%
%
\newpage

\centerline{\Large\bf Appendix}
\appendix

\section{Matrix integrals and measure factors \label{meas}}

In the main text, we have to calculate the matrix integration over the Lie groups $U(N)$, $SO(2N)$, $Sp(2N)$, $SO(2N+1)$. Here, we present 
a physicist's derivation of the measure factors using Fadeev-Popov determinant of path integral. The computation for $U(N)$ was described in \cite{Aharony:2003sx}. Suppose we compactify the theory on $S^1 \times S^2$, where the Euclidean time runs along the thermal circle $S^1$ with radius $\beta$. By integrating out all but the zero mode $\alpha$ of the time component of the gauge field, we obtain the following effective action for $\alpha$:
\be
 e^{-S_{\rm eff} (\alpha)} = \int dA \Delta_1 \exp \left(-S (A, \alpha) \right) \,,
\ee
where $\Delta_1$ is the Fadeev-Popov determinant corresponding to the gauge condition $\p_i A^i = 0$. The whole partition function is given by
\be
 Z = \int d \alpha \Delta_2 \exp \left(-S_{\rm eff} (\alpha) \right) \,,
\ee
where $\Delta_2 = \det' (\p_0 - i [\alpha, \cdot])$ is the other Fadeev-Popov determinant coming from the gauge fixing condition $\p_t \alpha = 0$. The integration measure of our interest in is $d\alpha \Delta_2$. It can be easily computed to give
\be
 \Delta_2 = \prod_{n \neq 0} \prod_{\alpha_i} \left( \frac{2\pi i n}{\beta} - i \lambda_i \alpha_i \right) \,,
\ee
where $\alpha_i$'s are root vectors of the Lie algebra. In the case of $SO(2N)$, the roots are given by $e_i \pm e_j$, so that
\be
 \Delta_2 &=& \prod_{n \neq 0} \prod_{i, j} \left( \frac{2\pi i n}{\beta} - i (\lambda_i - \lambda_j) \right)
                  \left( \frac{2\pi i n}{\beta} - i (\lambda_i + \lambda_j) \right) \\
 &=& \prod_{n \neq 0}\left( \frac{2\pi i n}{\beta} \right) \prod_{i < j} 
 \frac{4^2}{\beta^4 (\lambda_i - \lambda_j)^2 (\lambda_i + \lambda_j)^2} 
 \sin^2 \left( \frac{\beta(\lambda_i - \lambda_j)}{2} \right) \sin^2 \left( \frac{\beta(\lambda_i + \lambda_j)}{2} \right) \,.
 \nonumber
\ee
Therefore, 
\be
 d \alpha \Delta_2 = \prod_i d\lambda_i \prod_{i<j} \sin^2 \left( \frac{\lambda_i - \lambda_j}{2} \right) \sin^2 \left( \frac{\lambda_i + \lambda_j}{2} \right) \,.
\ee
Here, we redefined the $\beta \lambda \to \lambda$. If we write $U = e^{i \beta \alpha}$, this is the measure $DU$ of the matrix integral of $SO(2N)$, so we can write 
\be
 Z = \int DU \exp(-S_{\rm eff} (U)) \,.
\ee
Using the same technique, we can obtain the measure factors 
for other matrix integrals:
\be
 U(N) &:& \prod_i d\lambda_i \prod_{i<j} \sin^2 \left( \frac{\lambda_i - \lambda_j}{2} \right)  \,, \\
 SO(2N) &:&\prod_i d\lambda_i \prod_{i<j} \sin^2 \left( \frac{\lambda_i - \lambda_j}{2} \right) \sin^2 \left( \frac{\lambda_i + \lambda_j}{2} \right) \,, \\
 SO(2N+1) &:&  \prod_i d\lambda_i \prod_{i<j} \sin^2 \left( \frac{\lambda_i - \lambda_j}{2} \right) \sin^2 \left( \frac{\lambda_i + \lambda_j}{2} \right) \prod_k \sin^2 \left( \frac{\lambda_k}{2} \right) \,,  \\
Sp(2N) &:& \prod_i d\lambda_i \prod_{i<j} \sin^2 \left( \frac{\lambda_i - \lambda_j}{2} \right) \sin^2 \left( \frac{\lambda_i + \lambda_j}{2} \right) \prod_k \sin^2  \lambda_k \,.
\ee

\section{Character formulas for $SO(2r+1)$ and $SO(2r)$ \label{char-so}}

Define the character of a representation to be
\be\label{character}
\chi (h, t) = \Tr_{\{h_i\}} \exp(t_iH_i) \,,
\ee 
where $\{h_i\}$ are the highest weights labeling 
the representation, $\{t_i\}$ are real variables, and $\{H_i\}$ are the Cartan generators of our group in the standard basis. 
The sum over $i$ is taken in the RHS of the formula above. 
The $t$-variables here are related to the $y$-variables 
in the main text by $y=e^t$.

\paragraph{$SO(2r+1)$ :}

The highest weights of $SO(2r+1)$ satisfy 
$h_1 \geq h_2 \geq \cdots \geq h_r\geq 0$.
The Weyl character formula for $SO(2r+1)$ is 
\be\label{chara}
\chi (h, t)= \frac{\det[ \sinh(t_i(h_j+(r-j)+\half))] }{
\det[ \sinh(t_i( (r-j) +\half))] },
\ee
where one takes the determinant of the $(r\times r)$ matrix whose $(i,j)$ component
is written in the square bracket. One may check explicitly that the formula
reproduces the correct results for vector and spinor representations
\be
\chi ({\rm vector}) = 1 + 2 \sum_{i=1}^r \cosh(t_i) \,, 
\;\;\;\;\;
\chi ({\rm spinor}) = 2^r \prod_{i=1}^r \cosh \left(\frac{t_i}{2} \right)\,. 
\ee

\paragraph{$SO(2r)$ :}
The highest weights of $SO(2r)$ obey the condition 
$h_1 \geq h_2 \geq \cdots \geq h_{r-1} \geq |h_r|\geq 0$ 
($h_r$ may be negative).
The Weyl character formula for $SO(2r)$ is 
\be\label{charc}
\chi (h, t) = 
\frac{\det[ \sinh(t_i(h_j+r-j))] + \det[ \cosh(t_i(h_j+r-j))]
}{ \det[ \cosh(t_i( r-j))] } \,.
\ee
The vector and the two spinor characters may be worked out 
from this formula. 

\vskip 2cm

%

\end{document}